\documentclass[12pt]{article}
\pdfoutput=1

\usepackage{amsmath}
\usepackage{amssymb}
\usepackage{epsfig}
\usepackage{epstopdf}
\usepackage{latexsym}
\usepackage{color}
\numberwithin{equation}{section}
\usepackage[
      colorlinks=false,
      linkcolor=darkblue,  
      urlcolor=blue,    
      filecolor=blue,     
      citecolor=red,
linktocpage=true,
      pdfstartview=FitV,
      bookmarksopen=true    
      ]{hyperref}

\begin{document}

\begin{titlepage}

\centerline{\Huge \rm Magnetically-charged supersymmetric flows} 
\bigskip
\centerline{\Huge \rm of gauged $\mathcal{N}\,=\,8$ supergravity in five dimensions}
\bigskip
\bigskip
\bigskip
\bigskip
\bigskip
\bigskip
\centerline{\rm Minwoo Suh}
\bigskip
\centerline{\it Department of Physics, Kyungpook National University, Daegu 41566, Korea}
\bigskip
\centerline{\tt minwoosuh1@gmail.com} 
\bigskip
\bigskip
\bigskip
\bigskip
\bigskip
\bigskip
\bigskip

\begin{abstract}
\noindent We study magnetically-charged supersymmetric flow equations in a consistent truncation of gauged $\mathcal{N}\,=\,8$ supergravity in five dimensions. This truncation gives gauged $\mathcal{N}\,=\,2$ supergravity coupled to two vector multiplets and two hypermultiplets. We derive magnetically-charged flow equations of scalar fields from vector and hypermultiplets. It turns out that there could be only up to two nontrivial scalar fields out of eight from the hypermultiplets. Along the way, we recover some known $AdS_3$ solutions of the flow equations.
\end{abstract}

\bigskip
\bigskip
\bigskip
\vskip 7cm

\flushleft {April, 2018}

\end{titlepage}

\tableofcontents

\section{Introduction}

Supergravity has provided a simple and important venue in that the AdS/CFT correspondence \cite{Maldacena:1997re} could be tested in concrete examples. One of many notable examples is the so-called STU model. There are STU models in four- and five-dimensional gauged supergravity theories. To be specific, in five dimensions, we have an STU model in gauged $\mathcal{N}\,=\,2$ supergravity coupled to two vector multiplets. When we turn on a magnetic field, as it breaks the Lorentz invariance of the background, solutions cannot flow to $AdS_5$, but to $AdS_3{\times}\Sigma_k$ where $\Sigma_k$ is a two-dimensional surface with constant curvature. A number of such black string solutions are known in this model \cite{Chamseddine:1999xk, Klemm:2000nj, Cacciatori:2003kv, D'Hoker:2009mm, Almuhairi:2010rb, Almuhairi:2011ws, Donos:2011pn}.

Moreover, magnetically-charged flow solutions of the five-dimensional STU model enabled us to study renormalization group flows across dimensions holographically. Maldacena and Nunez found solutions interpolating $AdS_5$ and $AdS_3{\times}H^2$ which are dual to RG flows flowing from 4d $\mathcal{N}=4$ super Yang-Mills theory to 2d superconformal field theories \cite{Maldacena:2000mw}. Then, various wrapped brane solutions asymptote to $AdS_3{\times}S^2$ and $AdS_3{\times}H^2$ were found in \cite{Naka:2002jz, Cucu:2003bm, Cucu:2003yk}. These solutions were recently employed to holographically test $c$-extremization of 2d superconformal field theories \cite{Benini:2012cz, Benini:2013cda}. Holographic test of $c$-extremization was also considered from the aspect of three-dimensional gauged supergravity in \cite{Karndumri:2013iqa}. See also \cite{Amariti:2016mnz, Benini:2015bwz} for related works. 

So far, we have briefly reviewed achievements from studying the five-dimensional STU model which is gauged $\mathcal{N}\,=\,2$ supergravity coupled to two vector multiplets. One shortcoming of the STU model is that this model has only one critical point which is maximally supersymmetric and dual to 4d $\mathcal{N}=4$ super Yang-Mills theory. In order to study other critical points dual to various 4d superconformal field theories, one has to include other multiplets to the STU model, $e.g.$ vector or hypermultiplets. Indeed, in \cite{Bobev:2014jva}, magnetically-charged flows interpolating $AdS_5$ and $AdS_3{\times}H^2$ were obtained with vector and hypermultiplets. However, only one scalar field out of eight from hypermultiplets is nontrivial there. In this paper, we try to generalize their work and turn on four or less non-trivial scalar fields from hypermultiplets in the gauged $\mathcal{N}\,=\,8$ supergravity setting. As a main and final result, we will show that only up to two nontrivial scalar fields from hypermultiplets can be turned on. Along the way, we rediscover a number of $AdS_3{\times}H^2$ solutions found in \cite{Maldacena:2000mw, Benini:2012cz, Benini:2013cda}. Moreover, in \cite{Klemm:2016kxw}, the generic flow equations with vector and hypermultiplets were obtained. Our work could be also seen as a particular and concrete realization of their flow equations.

In section 2 we review a consistent truncation of gauged $\mathcal{N}\,=\,8$ supergravity. In section 3, as a warm up exercise, we rederive magnetically-charged flow equations with scalar fields only from vector multiplets. We recover some known solutions of the flow equations. In section 4, finally, we derive magnetically-charged flow equations with scalar fields from vector multiplets and also hypermultiplets. In section 5 concluding remarks are offered. In appendix A we briefly review gauged $\mathcal{N}\,=\,8$ supergravity in five dimensions. In appendix B some components of the supersymmetry variation for spin-1/2 fields are presented. In appendix C we present the equations of motion for the consistent truncation.

\section {A consistent truncation}

\subsection{The truncation}

We consider a consistent truncation of gauged $\mathcal{N}$ = 8 supergravity in five dimensions, \cite{Pernici:1985ju,Gunaydin:1984qu,Gunaydin:1985cu}, previously studied in \cite{Khavaev:2000gb} and more recently in \cite{Bobev:2010de}.

The 42 scalar fields of gauged $\mathcal{N}$ = 8 supergravity in five dimensions live on the coset manifold $E_{6(6)}/USp(8)$. The basic structure of the coset manifold explained in \cite{Gunaydin:1985cu} is summarized in appendix A. Fundamental representation of $E_{6(6)}$ is real and 27-dimensional. The infinitesimal $E_{6(6)}$ transformation in the $SL(6,\mathbb{R})$${\times}$$SL(2,\mathbb{R})$ basis, ($z_{IJ}$, $z^{I{\alpha}}$), is \cite{Gunaydin:1985cu}
\begin{align} \label{zls1}
{\delta}z_{IJ}\,=\,&-\Lambda^K\,_I\,z_{KJ}\,-\Lambda^K\,_J\,z_{IK}\,+\Sigma_{IJK\beta}\,z^{K\beta}\,, \\ \label{zls2}
{\delta}z_{I{\alpha}}\,=\,&\,\,\Lambda^I\,_K\,z^{K\alpha}\,+\,\Lambda^{\alpha}\,_{\beta}\,z_{I{\beta}}\,+\,\Sigma^{KLI\beta}\,z_{KL}\,,
\end{align}
where $\Lambda^I\,_J$ and $\Lambda^\alpha\,_\beta$ are real and traceless generators of $SL(6,\mathbb{R})$ and $SL(2,\mathbb{R})$ respectively, and $\Sigma_{IJK{\alpha}}$ is real and antisymmetric in $IJK$.

The truncation of our interest is obtained by a $\mathbb{Z}_4$-invariant subtruncation of $\mathbb{Z}_2{\times}\mathbb{Z}_2$-invariant sector \cite{Khavaev:2000gb, Bobev:2010de},{\footnote{This truncation is also obtained as a $U(1)_F$-invariant subtruncation of $U(1)_R$-invariant sector in \cite{Bobev:2014jva}.}} where the two $\mathbb{Z}_2$ generators in $SO(6)$ are
\begin{equation}
diag(-1,\,-1,\,-1,\,-1,\,1,\,1)\,, \qquad diag(1,\,1,\,-1,\,-1,\,-1,\,-1)\,.
\end{equation}
We have coset generators, $\Sigma_{IJK\alpha}$,
\begin{equation}
\Sigma\,=\,\frac{1}{12}\,\Sigma_{IJK\alpha}\,dx^I\,\wedge\,dx^J\,\wedge\,dx^K\,\wedge\,dy^\alpha\,.
\end{equation} 
With the complex coordinates, $z_1\,=\,x^1\,+\,i\,x^2$, $z_2\,=\,x^3\,-\,i\,x^4$, $z_3\,=\,x^5\,-\,i\,x^6$, and $z_4\,=\,y^1\,-\,i\,y^2$, {\footnote{We have $z_4\,=\,y^1\,-\,i\,y^2$ which is different from $z_4\,=\,y^1\,+\,i\,y^2$ in \cite{Khavaev:2000gb}}} we have
\begin{align} \label{param}
\Sigma_+\,=\,-\frac{1}{2}\sum_{i=1}^{4}\,Re[\varphi_i{e}^{i\theta_i}]\,\left(\Upsilon_i\,+\,\overline{\Upsilon}_i\right)\,, \notag \\
\Sigma_-\,=\,-\frac{1}{2i}\sum_{i=1}^{4}\,Im[\varphi_i{e}^{i\theta_i}]\,\left(\Upsilon_i\,-\,\overline{\Upsilon}_i\right)\,,
\end{align}
where
\begin{align}
\Upsilon_1\,=\,-\,dz_1\,\wedge\,dz_2\,\wedge\,dz_3\,\wedge\,dz_4\,, \qquad \Upsilon_2\,=\,-\,d\overline{z}_1\,\wedge\,d\overline{z}_2\,\wedge\,dz_3\,\wedge\,dz_4\,, \notag \\ \Upsilon_3\,=\,-\,d\overline{z}_1\,\wedge\,dz_2\,\wedge\,d\overline{z}_3\,\wedge\,dz_4\,, \qquad \Upsilon_4\,=\,-\,dz_1\,\wedge\,d\overline{z}_2\,\wedge\,d\overline{z}_3\,\wedge\,dz_4\,.
\end{align}
We also have two $SL(6, \mathbb{R})$ generators,{\footnote {As in \cite{Bobev:2010de}, we are reversing the sign $\alpha\,\rightarrow\,-\alpha$ compared to the conventions of \cite{Khavaev:2000gb}, to match earlier works, such as \cite{Freedman:1999gp}.}}
\begin{equation}
\Lambda_I\,^J\,=\,diag(\alpha\,+\,\beta\,,\,\alpha\,+\,\beta\,,\,\alpha\,-\,\beta\,,\,\alpha\,-\,\beta\,,\,-2\alpha\,,\,-2\alpha)\,.
\end{equation}
There are total ten real scalar fields, and they parametrize the coset manifold,
\begin{equation}
\mathcal{M}\,=\,\left(\frac{SU(1,1)}{U(1)}\right)^4\,\times\,SO(1,1)^2\,.
\end{equation}
This truncation gives gauged $\mathcal{N}\,=\,2$ supergravity in five dimensions coupled to two vector multiplets and two hypermultiplets. The scalar fields, $\alpha$ and $\beta$, parametrizing the $SO(1,1)$ factors, belong to two $\mathcal{N}\,=\,2$ vector multiplets. The scalar fields, $\varphi_j$ and $\theta_j$, $j\,=\,1,\ldots, 4$, parametrizing the $SU(1,1)/U(1)$ factors, belong to two $\mathcal{N}\,=\,2$ hypermultiplets.{\footnote {It is natural and sometimes convenient to employ a parametrization of complex scalar fields, $\zeta_j\,=\,\tanh{\varphi_j}e^{i\theta_j}$.}}

The Lagrangian of the truncation is given by
\begin{align} \label{lag}
e^{-1}\,\mathcal{L}\,=\,&-\,\frac{1}{4}\,R\,-\,\frac{1}{4}\,\Big[e^{4\alpha-4\beta}F_{\mu\nu}^{(1)}\,F^{(1)\mu\nu}\,+\,e^{4\alpha+4\beta}\,F_{\mu\nu}^{(2)}\,F^{(2)\mu\nu}\,+\,e^{-8\alpha}\,F_{\mu\nu}^{(3)}\,F^{(3)\mu\nu}\,\Big]\, \notag \\
&+\,\frac{1}{2}\sum_{j=1}^4\,(\partial_\mu\varphi_j)^2\,+\,3\,(\partial_\mu\alpha)^2\,+\,(\partial_\mu\beta)^2\, \notag \\
&+\,\frac{1}{8}\,\sinh^2(2\varphi_1)\,\left(\partial_\mu\theta_1\,+\,(A_\mu^{(1)}\,+\,A_\mu^{(2)}\,-\,A_\mu^{(3)})\right)^2\, \notag \\
&+\,\frac{1}{8}\,\sinh^2(2\varphi_2)\,\left(\partial_\mu\theta_2\,-\,(A_\mu^{(1)}\,-\,A_\mu^{(2)}\,+\,A_\mu^{(3)})\right)^2\, \notag \\
&+\,\frac{1}{8}\,\sinh^2(2\varphi_3)\,\left(\partial_\mu\theta_3\,+\,(-A_\mu^{(1)}\,+\,A_\mu^{(2)}\,+\,A_\mu^{(3)})\right)^2\, \notag \\
&+\,\frac{1}{8}\,\sinh^2(2\varphi_4)\,\left(\partial_\mu\theta_4\,-\,(A_\mu^{(1)}\,+\,A_\mu^{(2)}\,+\,A_\mu^{(3)})\right)^2\,-\,\mathcal{P}\,.
\end{align}
There are three vector fields: two are from two $\mathcal{N}\,=\,2$ vector multiplets and one is the graviphoton from an $\mathcal{N}\,=\,2$ gravity multiplet. The scalar potential is
\begin{equation}
\mathcal{P}\,=\,\frac{g^2}{8}\,\left[\sum_{j=1}^4\,\left(\frac{\partial{W}}{\partial\varphi_j}\right)^2\,+\frac{1}{6}\,\left(\frac{\partial{W}}{\partial\alpha}\right)^2\,+\,\frac{1}{2}\,\left(\frac{\partial{W}}{\partial\beta}\right)^2\,\right]\,-\,\frac{g^2}{3}\,W^2\,,
\end{equation}
where the superpotential is
\begin{align} \label{sp}
W\,=\,-\frac{1}{4}\Big[&\left(e^{-2\alpha+2\beta}+e^{-2\alpha-2\beta}-e^{4\alpha}\right)\cosh(2\varphi_1)+\left(e^{-2\alpha+2\beta}-e^{-2\alpha-2\beta}+e^{4\alpha}\right)\cosh(2\varphi_2) \notag \\ 
+&\left(-e^{-2\alpha+2\beta}+e^{-2\alpha-2\beta}+e^{4\alpha}\right)\cosh(2\varphi_3)+\left(e^{-2\alpha+2\beta}+e^{-2\alpha-2\beta}+e^{4\alpha}\right)\cosh(2\varphi_4)\Big]\,.
\end{align}
The scalar potential is independent of the phases, $\theta_j$. Refer to \cite{Khavaev:2000gb, Bobev:2010de} for the critical points of the scalar potential and more details of the truncation.

\subsection{The supersymmetry equations}

For magnetically-charged flows, we consider the background,
\begin{equation} \label{metric}
ds^2\,=\,e^{2U(r)}\,\left(dt^2\,-\,dz^2\right)\,-\,e^{2V(r)}\,\left(dx^2\,+\,f_k^2(x)\,dy^2\right)\,-\,\frac{dr^2}{p^2(r)}\,,
\end{equation}
where
\begin{equation}
 f_0\,=\,1\,, \qquad f_+\,=\,\sin{x}\,, \qquad f_-\,=\,\sinh{x}\,,
\end{equation}
and $\Sigma_k\,=\,\mathbb{R}^2,\,S^2,\,H^2$, for $k\,=\,0,\,+1,\,-1$, respectively, for the $x$-$y$ plane. The only non-zero component of the gauge field, $A_{\mu{IJ}}$, is given by
\begin{equation}
A_{yIJ}=\left(
\begin{array}{llllll}
 0 & q_1 & 0 & 0 & 0 & 0 \\
 -q_1 & 0 & 0 & 0 & 0 & 0 \\
 0 & 0 & 0 & q_2 & 0 & 0 \\
 0 & 0 & -q_2 & 0 & 0 & 0 \\
 0 & 0 & 0 & 0 & 0 & q_3 \\
 0 & 0 & 0 & 0 & -q_3 & 0
\end{array}
\right)\,\int{f}_k(x)\,dx\,.
\end{equation}
Hence, the field strength is
\begin{equation}
F_{xyIJ}=\left(
\begin{array}{llllll}
 0 & q_1 & 0 & 0 & 0 & 0 \\
 -q_1 & 0 & 0 & 0 & 0 & 0 \\
 0 & 0 & 0 & q_2 & 0 & 0 \\
 0 & 0 & -q_2 & 0 & 0 & 0 \\
 0 & 0 & 0 & 0 & 0 & q_3 \\
 0 & 0 & 0 & 0 & -q_3 & 0
\end{array}
\right)\,f_k(x)\,.
\end{equation}

We consider the superpotential and the spinors in five dimensions. The superpotential, $W$, is obtained as one of the eigenvalues of $W_{ab}$ tensor \cite{Freedman:1999gp},
\begin{equation}
W_{ab}\,\eta^b_{(k)}\,=\,W\,\eta^a_{(k)}\,,
\end{equation}
where $k\,=\,1,\,2$ and $W$ is given in \eqref{sp}. The eigenvectors, $\eta^a_{(1)}$, $\eta^a_{(2)}$, are
\begin{align}
\eta^a_{(1)}\,&=\,(0,\,1,\,0,\,1,\,1,\,0,\,-1,\,0)\,, \\
\eta^a_{(2)}\,&=\,(-1,\,0,\,1,\,0,\,0,\,1,\,0,\,1)\,,
\end{align}
and they are related to each other by
\begin{equation}
\Omega_{ab}\,\eta^b_{(1)}\,=\,-\,\eta^a_{(2)},\,\,\,\,\,\,\,\,\,\, \Omega_{ab}\,\eta^b_{(2)}\,=\,+\,\eta^a_{(1)}\,,
\end{equation}
where $\Omega_{ab}$ is the $USp(8)$ symplectic form in {\it e.g.} \cite{Freedman:1999gp}. We employ the gamma matrix conventions in \cite{Freedman:1999gp}. The spinors are
\begin{align}
\epsilon^a\,&=\,\eta^a_{(1)}\,\hat{\epsilon}_1\,+\,\eta^a_{(2)}\,\hat{\epsilon}_2\,, \\
\epsilon_a\,&=\,\Omega_{ab}\,\epsilon^b\,=\,-\,\eta^a_{(2)}\,\hat{\epsilon}_1\,+\,\eta^a_{(1)}\,\hat{\epsilon}_2\,,
\end{align}
where $\hat{\epsilon}_1$ and $\hat{\epsilon}_2$ are spinors with four complex components.

The supersymmetry equations are obtained by setting the supersymmetry variations of fermionic fields, {\it i.e.} the spin-3/2 and spin-1/2 fields, to zero. The bosonic parts of the variations are \cite{Gunaydin:1985cu}
\begin{align} \label{3/2A}
\delta\,\psi_{\mu{a}}\,&=\,D_\mu\,\epsilon_a\,-\,\frac{1}{6}\,g\,W_{ab}\,\gamma_\mu\,\epsilon^b\,-\frac{1}{6}\,H_{\nu\rho{ab}}\,(\gamma^{\nu\rho}\,\gamma_\mu\,+2\,\gamma^\nu\,\delta^\rho_\mu)\,\epsilon^b\,, \\ \label{1/2A}
\delta\,\chi_{abc}\,&=\,\sqrt{2}\,\left[\gamma^\mu\,P_{\mu{abcd}}\,\epsilon^d\,-\,\frac{1}{2}\,g\,A_{dabc}\,\epsilon^d\,-\frac{3}{4}\,\gamma^{\mu\nu}\,H_{\mu\nu[ab}\,\epsilon_{c]|}\right]\,,
\end{align}
where
\begin{equation}
D_\mu\,\epsilon_a\,=\,\partial_\mu\,\epsilon_a\,+\frac{1}{4}\,\omega_{\mu{ij}}\,\gamma^{ij}\,\epsilon_a\,+\,Q_{\mu{a}}\,^b\,\epsilon_b\,.
\end{equation}
We define
\begin{equation}
H_{\mu\nu}\,^{ab}\,=\,F_{\mu\nu}\,^{ab}\,+\,B_{\mu\nu}\,^{ab}\,,
\end{equation}
where
\begin{equation}
F_{\mu\nu}\,^{ab}\,=\,F_{\mu\nu{IJ}}\,\mathcal{V}^{IJab}\,, \,\,\,\,\,\,\,\,\,\, B_{\mu\nu}\,^{ab}\,=\,B_{\mu\nu}\,^{I\alpha}\,\mathcal{V}_{I\alpha}\,^{ab}\,.
\end{equation}
For the consistent truncation we have, we have
\begin{equation}
F_{\mu\nu{IJ}}\,=\,\partial_\mu\,A_{\nu{IJ}}\,-\,\partial_\nu\,A_{\mu{IJ}}\,, \,\,\,\,\,\,\,\,\,\, B_{\mu\nu}\,^{I\alpha}\,=\,0\,.
\end{equation}
Hence, the only non-zero component is
\begin{equation}
H_{xy\,{ab}}\,=\,F_{xyIJ}\,\mathcal{V}^{IJab}\,.
\end{equation}

It is convenient to define the quantities, $H$, $Q$, along with the superpotential, $W$,
\begin{align}
W_{ab}\,\eta^b_{(1)}\,&\,=\,W\,\eta^a_{(1)}\,, \,\,\,\,\,\,\,\,\,\, W_{ab}\,\eta^b_{(2)}\,=\,W\,\eta^a_{(2)}\,, \\
H_{xy\,{ab}}\,\eta^b_{(1)}\,&=\,-\,H\,\eta^a_{(2)}\,, \,\,\,\,\,\,\,\,\,\, H_{xy\,{ab}}\,\eta^b_{(2)}\,=\,+\,H\,\eta^a_{(1)}\,, \\
Q_{y\,{a}}\,^b\,\eta^b_{(1)}\,&=\,-\,Q\,\eta^a_{(2)}\,, \,\,\,\,\,\,\,\,\,\, Q_{y\,{a}}\,^b\,\eta^b_{(2)}\,=\,+\,Q\,\eta^a_{(1)}\,.
\end{align}
Hence, we have
\begin{align}
W_{ab}\,\epsilon^b\,&=\,W\,\epsilon^a\,, \\
H_{xy\,{ab}}\,\epsilon^b\,&=\,H_{xy\,{ab}}\,(\eta^b_{(1)}\,\hat{\epsilon}_1\,+\,\eta^b_{(2)}\,\hat{\epsilon}_2)\,=\,-H\,\eta^a_{(2)}\,\hat{\epsilon}_1\,+\,H\,\eta^a_{(1)}\,\hat{\epsilon}_2\,=\,H\,\epsilon_a\,, \\
Q_{y\,{a}}\,^b\,\epsilon_b\,&=\,Q_{y\,{a}}\,^b\,(-\,\eta^b_{(2)}\,\hat{\epsilon}_1\,+\,\eta^b_{(1)}\,\hat{\epsilon}_2)\,=\,-Q\,\eta^a_{(1)}\,\hat{\epsilon}_1\,-Q\,\eta^a_{(2)}\,\hat{\epsilon}_2\,=\,-\,Q\,\epsilon^a\,.
\end{align}

\section{Magnetically-charged flow equations: vector multiplets}

In the consistent truncation we have, there are ten scalar fields, $\alpha$, $\beta$ from $\mathcal{N}\,=\,2$ vector multiplets, and $\varphi_j$, $\theta_j$, $j\,=\,1,\ldots, 4$ from $\mathcal{N}\,=\,2$ hypermultiplets. In this section, as a warm up exercise and to set up the conventions, we rederive magnetically-charged flow equations only with the scalar fields from vector multiplets. Similar equations were obtained in gauged $\mathcal{N}\,=\,2$ supergravity coupled to vector multiplets in five dimensions in \cite{Chamseddine:1999xk, Klemm:2000nj, Cacciatori:2003kv} and \cite{Almuhairi:2011ws, Donos:2011pn}. In the next section, we will consider inclusion of the scalar fields from hypermultiplets. 

\subsection{The flow equations}

In this section, we only turn on the scalar fields from vector multiplets, $\alpha$ and $\beta$. We set the scalar fields from hypermultiplets, $\varphi_j\,=\,0$, $\theta_j\,=\,0$, $j\,=\,1,\ldots, 4$. The superpotential in \eqref{sp} reduces to
\begin{equation}
W=-\frac{1}{2}\left(e^{-2\alpha+2\beta}+e^{-2\alpha-2\beta}+e^{4\alpha}\right)\,.
\end{equation}
This superpotential does not have any critical point beside the maximally supersymmetric one.

Now we consider the spin-3/2 field variations in \eqref{3/2A}. For $\mu\,=\,t,\,x,\,y,\,r$, respectively, we obtain
\begin{equation} \label{sv1}
p\,U'\,\gamma^4\,\epsilon_a\,-\,\frac{1}{3}\,g\,W\,\epsilon^a\,-\,F\,e^{-2V}\,\gamma^1\,\gamma^2\,\epsilon_a\,=\,0\,, 
\end{equation}
\begin{equation} \label{sv2}
p\,V'\,\gamma^4\,\epsilon_a\,-\,\frac{1}{3}\,g\,W\,\epsilon^a\,+\,2\,F\,e^{-2V}\,\gamma^1\,\gamma^2\,\epsilon_a\,=\,0\,,
\end{equation}
\begin{equation} \label{sv3}
p\,V'\,\gamma^4\,\epsilon_a\,-\,\frac{1}{3}\,g\,W\,\epsilon^a\,+\,2\,F\,e^{-2V}\,\gamma^1\,\gamma^2\,\epsilon_a\,+\,\frac{e^{-V}}{f_k}\,\gamma^1\,\left(\partial_x{f}_k\,\epsilon_a\,+\,2\,Q\,\gamma^1\,\gamma^2\,\epsilon^a\right)\,=\,0\,,
\end{equation}
\begin{equation} \label{sv4}
2\,p\,\gamma^4\,\partial_r\,\epsilon_a\,-\,\frac{1}{3}\,g\,W\,\epsilon^a\,-\,F\,e^{-2V}\,\gamma^1\,\gamma^2\,\epsilon_a\,=\,0\,, 
\end{equation}
where 
\begin{align}
Q\,=&\,-\,\frac{g}{2}\,(q_1\,-\,q_2\,+\,q_3)\,\int{f}_k\,dx\,, \\
F\,=&\,\frac{1}{3}\left(q_1e^{2\alpha-2\beta}-q_2e^{2\alpha+2\beta}+q_3e^{-4\alpha}\right)\,,
\end{align}
and the prime denotes the derivative with respective to $r$. The equation for $\mu\,=\,z$ is identical to the one for $\mu\,=\,t$. We employ the projection conditions,
\begin{equation} \label{proj}
\gamma^4\,\epsilon_a\,=\,-\,\epsilon^a\,, \qquad \gamma^1\,\gamma^2\,\epsilon_a\,=\,+\,\epsilon^a\,.
\end{equation}
From \eqref{sv1} and \eqref{sv2}, we obtain
\begin{equation} \label{fone}
p\,U'\,+\,\frac{1}{3}\,g\,W\,+\,F\,e^{-2V}\,=\,0\,, \qquad p\,V'\,+\,\frac{1}{3}\,g\,W\,-\,2\,F\,e^{-2V}\,=\,0\,.
\end{equation}
The difference of \eqref{sv2} and \eqref{sv3} gives
\begin{equation} \label{precond}
\epsilon_a\,\pm\,g\,(q_1\,-\,q_2\,+\,q_3)\,\gamma^1\,\gamma^2\,\epsilon^a\,=\,0\,,
\end{equation}
and we obtain a condition,
\begin{equation} \label{scondition}
g\,(q_1\,-\,q_2\,+\,q_3)\,=\,k\,.
\end{equation}

Now we consider the spin-1/2 field variations in \eqref{1/2A}. They reduce to
\begin{align}
\gamma^4\,\alpha'\,\epsilon_a\,+\,\frac{g}{12}\,\frac{\partial{W}}{\partial\alpha}\,\epsilon^a\,+\,\gamma^1\,\gamma^2\,X_\alpha\,\epsilon_a\,=\,0\,, \notag \\
\gamma^4\,\beta'\,\epsilon_a\,+\,\frac{g}{4}\,\frac{\partial{W}}{\partial\beta}\,\epsilon^a\,+\,\gamma^1\,\gamma^2\,X_\beta\,\epsilon_a\,=\,0\,,
\end{align}
where
\begin{align}
X_\alpha\,=&\,-\,\frac{1}{6}\,e^{-2V}\,(q_1\,e^{2\alpha-2\beta}\,-\,q_2\,e^{2\alpha+2\beta}\,-\,2\,q_3\,e^{-4\alpha})\,, \notag \\
X_\beta\,=&\,+\,\frac{1}{2}\,e^{-2V}\,(q_1\,e^{2\alpha-2\beta}\,+\,q_2\,e^{2\alpha+2\beta})\,.
\end{align}
Employing the projection conditions, \eqref{proj}, we obtain
\begin{equation} \label{ftwo}
\alpha'\,-\,\frac{g}{12}\,\frac{\partial{W}}{\partial\alpha}\,-\,X_\alpha\,=\,0\,, \qquad \beta'\,-\,\frac{g}{4}\,\frac{\partial{W}}{\partial\beta}\,-\,X_\beta\,=\,0\,.
\end{equation}

Now let us summarize the supersymmetric flow equations obtained in \eqref{fone} and \eqref{ftwo},
\begin{equation} \label{susy1}
p\,U'\,=\,-\,\frac{1}{3}\,g\,W\,+\,X_U\,, \qquad p\,V'\,=\,-\,\frac{1}{3}\,g\,W\,+\,X_V\,. \notag
\end{equation}
\begin{equation}
\alpha'\,=\,\frac{g}{12}\,\frac{\partial{W}}{\partial\alpha}\,+\,X_\alpha\,, \qquad
\beta'\,=\,\frac{g}{4}\,\frac{\partial{W}}{\partial\beta}\,+\,X_\beta\,,
\end{equation}
where
\begin{align} \label{susy2}
X_U\,=&\,-\,\frac{1}{3}\,e^{-2V}\,\left(q_1\,e^{2\alpha-2\beta}\,-\,q_2\,e^{2\alpha+2\beta}\,+\,q_3\,e^{-4\alpha}\right)\,, \notag \\
X_V\,=&\,+\,\frac{2}{3}\,e^{-2V}\,\left(q_1\,e^{2\alpha-2\beta}\,-\,q_2\,e^{2\alpha+2\beta}\,+\,q_3\,e^{-4\alpha}\right)\,, \notag \\
X_\alpha\,=&\,-\,\frac{1}{6}\,e^{-2V}\,\left(q_1\,e^{2\alpha-2\beta}\,-\,q_2\,e^{2\alpha+2\beta}\,-\,2\,q_3\,e^{-4\alpha}\right)\,, \notag \\
X_\beta\,=&\,+\,\frac{1}{2}\,e^{-2V}\,\left(q_1\,e^{2\alpha-2\beta}\,+\,q_2\,e^{2\alpha+2\beta}\right)\,,
\end{align}
with a condition,
\begin{equation} \label{cond}
g\,(q_1\,-\,q_2\,+\,q_3)\,=\,k\,,
\end{equation}
and the $x$-$y$ plane is $\Sigma_k\,=\,\mathbb{R}^2,\,S^2,\,H^2$, for $k\,=\,0,\,+1,\,-1$, respectively.

\subsection{Some known solutions}

In this subsection, we recover some known analytic and numerical solutions of the supersymmetric flow equations obtained in \eqref{susy1} with \eqref{susy2} and \eqref{cond}. 

\subsubsection{Supersymmetric domain wall solutions of $AdS_5$ and $AdS_3\times\mathbb{R}^2$}

We numerically solve the supersymmetric flow equations, and obtain a magnetically-charged domain wall solution interpolating $AdS_5$ and $AdS_3\times\mathbb{R}^2$, which was previously studied in \cite{Donos:2011pn}.

In \cite{Almuhairi:2010rb, Almuhairi:2011ws}, an $AdS_3{\times}\mathbb{R}^2$ solution was found which is a solution of the supersymmetric flow equations. The solution is given by the metric,
\begin{equation} \label{ads3r2}
ds^2\,=\,e^{r/L}(dt^2-dz^2)-dr^2-(dx^2+dy^2)\,,
\end{equation}
and the scalar fields,
\begin{equation}
\alpha=\alpha_*\,, \qquad \beta=\beta_*\,,
\end{equation}
where $\alpha_*$ and $\beta_*$ are some constants. The constant scalar fields give the values of
\begin{equation}
q_1=-e^{\alpha_*+\beta_*}\,, \qquad q_2=e^{-\alpha_*-\beta_*}\,, \qquad q_3=e^{2\beta_*}\,,
\end{equation}
and
\begin{equation}
L=\frac{2}{q_1^2+q_2^2+q_3^2}\,,
\end{equation}
and they satisfy the condition, \eqref{cond}, with $k\,=\,0$.

Now we numerically solve the supersymmetric flow equations and recover the domain wall solution of \cite{Donos:2011pn}, that interpolate between an $AdS_5$ in the UV and the $AdS_3\times\mathbb{R}^2$ solution of \eqref{ads3r2} in the IR. We choose $g=2$, $\alpha_*=(\log2)/3$, $\beta_*=0$, and a numerical solution is depicted in Figure 1.
\begin{figure}[h!]
\begin{center}
\includegraphics[width=2.7in]{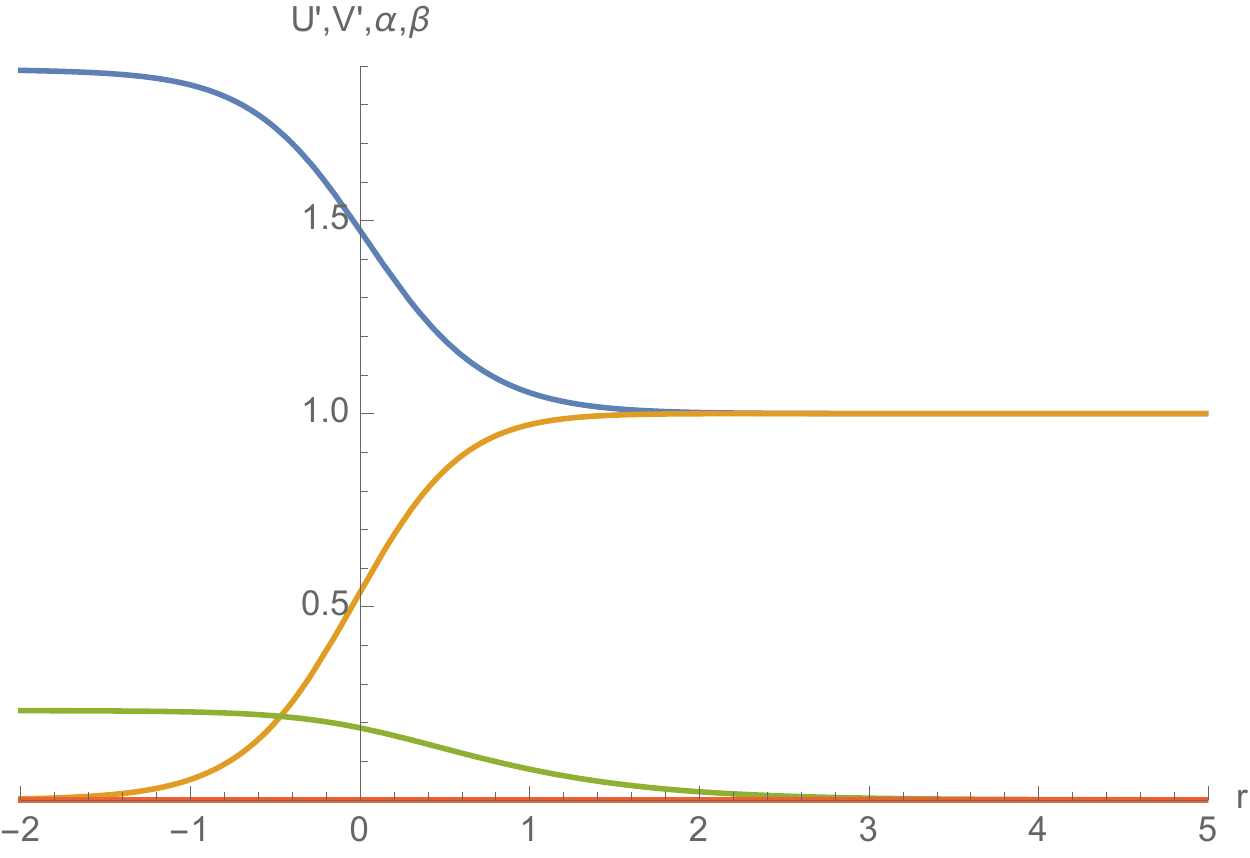}
\caption{\it A domain wall solution of $U'$ (blue), $V'$ (orange), $\alpha$ (green), and $\beta$ (red).}
\end{center}
\end{figure}

\subsubsection{Supersymmetric black string solutions with trivial scalar fields}

The Chamseddine-Sabra solution in (23) of \cite{Chamseddine:1999xk} and the Klemm-Sabra solution in (30) of \cite{Klemm:2000nj} satisfy the supersymmetric flow equations. The solutions are given by the metric,
\begin{equation}
ds^2\,=\,(l\,r)^{1/2}\,\left(k\,\frac{1}{3\,l\,r}\,+\,l\,r\right)^{3/2}\,\left(dt^2\,-\,dz^2\right)\,-\,r^2\,\left(dx^2\,+\,\sinh^2x\,dy^2\right)\,-\,\left(k\,\frac{1}{3\,l\,r}\,+\,l\,r\right)^{-2}\,dr^2\,,
\end{equation}
and trivial scalar fields,
\begin{equation}
\alpha\,=\,\beta\,=\,0\,,
\end{equation}
with a condition,
\begin{equation}
q_1\,=\,-\,q_2\,=\,q_3\,=\,\,\frac{k}{3g}\,,
\end{equation}
where $l\,=\,g/2$ and $k=+1,\,-1$ for $S^2$ and $H^2$, respectively. In the near horizon limit, the solution asymptotes to $AdS_3{\times}S^2$ \cite{Chamseddine:1999xk} or $AdS_3{\times}H^2$ \cite{Klemm:2000nj}. Asymptotically $AdS_3{\times}H^2$ solution is a black string solution with a regular horizon at $r=1/(\sqrt{3}l)$, but asymptotically $AdS_3{\times}S^2$ has a naked singularity.

\subsubsection{Supersymmetric black string solutions with nontrivial scalar fields}

The Cacciatori-Klemm-Sabra black string solution in (6.4) of \cite{Cacciatori:2003kv} satisfies the supersymmetric flow equations. The solution is given by the metric,
\begin{align}
ds^2\,=&\,e^{kr^2/6}\,r^{-2}\,\left(1\,-k\,\frac{r^2}{6}\right)^{-2/3}\,\left(dt^2\,-\,dz^2\right)\,-\,(l\,r)^{-2}\,\left(1\,-k\,\frac{r^2}{6}\right)^{4/3}\,\left(dr^2\,+\,dx^2\,+\,f_k^2(x)\,dy^2\right)\,,
\end{align}
and nontrivial scalar fields,
\begin{equation}
e^\alpha\,=\,e^{\beta/3}\,=\,\left(1\,-k\,\frac{r^2}{6}\right)^{1/{12}}\,,
\end{equation}
with a condition,
\begin{equation}
q_1\,=\,-\,q_2\,=\,q_3\,=\,\,\frac{k}{3g}\,,
\end{equation}
where $l\,=\,g/2$. In the near horizon limit, the solutions asymptote to $AdS_3{\times}S^2$ or $AdS_3{\times}H^2$ for $k=+1,\,-1$, respectively.

\section{Magnetically-charged flow equations: vector and hypermultiplets}

\subsection{The flow equations}

In the consistent truncation we have, there are ten scalar fields, $\alpha$, $\beta$ from $\mathcal{N}\,=\,2$ vector multiplets, and $\varphi_j$, $\theta_j$, $j\,=\,1,\ldots, 4$ from $\mathcal{N}\,=\,2$ hypermultiplets. In the previous section, we obtained the flow equations only with the scalar fields from $\mathcal{N}\,=\,2$ vector multiplets. In this section, we finally derive magnetically-charged flow equations with inclusion of the scalar fields from $\mathcal{N}\,=\,2$ hypermultiplets. 

In this section, we will fix $p(r)\,=\,1$ in the background metric of \eqref{metric}. We will assume that the scalar fields, $\theta_j$, $j\,=\,1,\ldots, 4$, are functions of the $y$-coordinate only, $\theta_j\,=\,\theta_j(y)$. The rest of the scalar fields and functions depend on the radial $r$-coordinate only. The reason will be clear below.

First we consider the spin-3/2 field variations in \eqref{3/2A}. For $\mu\,=\,t,\,x,\,y,\,r$, respectively, we obtain
\begin{equation} \label{ss1}
U'\,\gamma^4\,\epsilon_a\,-\,\frac{1}{3}\,g\,W\,\epsilon^a\,-\,F\,e^{-2V}\,\gamma^1\,\gamma^2\,\epsilon_a\,=\,0\,, 
\end{equation}
\begin{equation} \label{ss2}
V'\,\gamma^4\,\epsilon_a\,-\,\frac{1}{3}\,g\,W\,\epsilon^a\,+\,2\,F\,e^{-2V}\,\gamma^1\,\gamma^2\,\epsilon_a\,=\,0\,,
\end{equation}
\begin{equation} \label{ss3}
V'\,\gamma^4\,\epsilon_a\,-\,\frac{1}{3}\,g\,W\,\epsilon^a\,+\,2\,F\,e^{-2V}\,\gamma^1\,\gamma^2\,\epsilon_a\,+\,\frac{e^{-V}}{f_k}\,\gamma^1\,\left(\partial_x{f}_k\,\epsilon_a\,+\,2\,Q\,\gamma^1\,\gamma^2\,\epsilon^a\right)\,=\,0\,,
\end{equation}
\begin{equation} \label{ss4}
2\,\gamma^4\,\partial_r\,\epsilon_a\,-\,\frac{1}{3}\,g\,W\,\epsilon^a\,-\,F\,e^{-2V}\,\gamma^1\,\gamma^2\,\epsilon_a\,=\,0\,, 
\end{equation}
where
\begin{align}
Q\,=&\,\frac{g}{4}\Huge[(q_1-q_2-q_3)\cosh(2\varphi_1)+(q_1+q_2+q_3)\cosh(2\varphi_2) \notag \\
&+(-q_1-q_2+q_3)\cosh(2\varphi_3)+(q_1-q_2+q_3)\cosh(2\varphi_4)\Huge] \int{f}_kdx \notag \\
+&\frac{1}{2}\left[\sinh^2\varphi_1\dot{\theta}_1-\sinh^2\varphi_2\dot{\theta}_2+\sinh^2\varphi_3\dot{\theta}_3-\sinh^2\varphi_4\dot{\theta}_4\right]\,, \\
F\,=&\,\frac{1}{3}\left(q_1e^{2\alpha-2\beta}-q_2e^{2\alpha+2\beta}+q_3e^{-4\alpha}\right)\,,
\end{align}
and the prime and the dot denote the derivatives with respective to $r$ and $y$, respectively. The equation for $\mu\,=\,z$ is identical to the one for $\mu\,=\,t$. We employ the projection conditions,
\begin{equation} \label{proj2}
\gamma^4\,\epsilon_a\,=\,-\,\epsilon^a\,, \qquad \gamma^1\,\gamma^2\,\epsilon_a\,=\,+\,\epsilon^a\,.
\end{equation}
From \eqref{ss1} and \eqref{ss2}, we obtain
\begin{equation} \label{warp}
U'\,+\,\frac{1}{3}\,g\,W\,+\,F\,e^{-2V}\,=\,0\,, \qquad V'\,+\,\frac{1}{3}\,g\,W\,-\,2\,F\,e^{-2V}\,=\,0\,.
\end{equation}
The difference of \eqref{ss2} and \eqref{ss3} gives
\begin{align} \label{precond}
\frac{k}{g}\,=\,\frac{1}{2}&\Huge[(q_1-q_2-q_3)\cosh(2\varphi_1)+(q_1+q_2+q_3)\cosh(2\varphi_2) \notag \\
+&(-q_1-q_2+q_3)\cosh(2\varphi_3)+(q_1-q_2+q_3)\cosh(2\varphi_4)\Huge] \notag \\
+&\frac{1}{g\int{f}_kdx}\left[\sinh^2\varphi_1\dot{\theta}_1-\sinh^2\varphi_2\dot{\theta}_2+\sinh^2\varphi_3\dot{\theta}_3-\sinh^2\varphi_4\dot{\theta}_4\right]\,.
\end{align}

Now we consider the spin-1/2 field variations in \eqref{1/2A}. We present the details in appendix B. We employ the projection conditions, \eqref{proj2}, and the spin-1/2 field variations reduce to
\begin{equation} \label{s1}
\alpha'\,=\,\frac{g}{12}\frac{\partial{W}}{\partial\alpha}+X_{\alpha}\,,\qquad
\beta'\,=\,\frac{g}{4}\frac{\partial{W}}{\partial\beta}+X_{\beta}\,, \notag
\end{equation}
\begin{equation} 
\varphi_1'=\,\frac{g}{2}\frac{\partial{W}}{\partial\varphi_1}\,, \qquad
\varphi_2'\,=\,\frac{g}{2}\frac{\partial{W}}{\partial\varphi_2}\,, \qquad
\varphi_3'\,=\,\frac{g}{2}\frac{\partial{W}}{\partial\varphi_3}\,, \qquad
\varphi_4'\,=\,\frac{g}{2}\frac{\partial{W}}{\partial\varphi_4}\,, \notag
\end{equation}
\begin{align} \label{thetas}
\dot{\theta}_1\,=\,-g(q_1-q_2-q_3)\int{f}_kdx\,, \qquad \dot{\theta}_2\,=\,g(q_1+q_2+q_3)\int{f}_kdx\,, \notag \\ \dot{\theta}_3\,=\,-g(-q_1-q_2+q_3)\int{f}_kdx\,, \qquad \dot{\theta}_4\,=\,g(q_1-q_2+q_3)\int{f}_kdx\,.
\end{align}
Recall two equations from the spin-3/2 field variations in \eqref{warp},
\begin{equation} \label{s2}
U'\,=\,-\,\frac{1}{3}\,g\,W\,+\,X_U\,, \qquad V'\,=\,-\,\frac{1}{3}\,g\,W\,+\,X_V\,,
\end{equation}
and
\begin{align}
X_U\,=&\,-\,\frac{1}{3}\,e^{-2V}\,\left(q_1\,e^{2\alpha-2\beta}\,-\,q_2\,e^{2\alpha+2\beta}\,+\,q_3\,e^{-4\alpha}\right)\,, \notag \\
X_V\,=&\,+\,\frac{2}{3}\,e^{-2V}\,\left(q_1\,e^{2\alpha-2\beta}\,-\,q_2\,e^{2\alpha+2\beta}\,+\,q_3\,e^{-4\alpha}\right)\,, \notag \\
X_\alpha\,=&\,-\,\frac{1}{6}\,e^{-2V}\,\left(q_1\,e^{2\alpha-2\beta}\,-\,q_2\,e^{2\alpha+2\beta}\,-\,2\,q_3\,e^{-4\alpha}\right)\,, \notag \\
X_\beta\,=&\,+\,\frac{1}{2}\,e^{-2V}\,\left(q_1\,e^{2\alpha-2\beta}\,+\,q_2\,e^{2\alpha+2\beta}\right)\,.
\end{align}
These are the complete supersymmetric flow equations. However, we observe that the scalar field, $\theta_j$, equations, in \eqref{thetas}, are {\it contradictory} by themselves, as the left-hand sides are functions of $y$ only, but the right-hand sides are functions of $x$ only. Therefore, these equations reduce to a number of constraints on the scalar fields and the gauge fields, as we will look in detail in the following subsections.{\footnote{We are very grateful to an anonymous referee who pointed this to us.}}

\subsubsection{Flow equations with one nontrivial scalar field from hypermultiplets}

If we turn off a scalar field, $\varphi_j$, as we observe from \eqref{param}, it automatically turns off the scalar field, $\theta_j$, from the truncation. In this subsection, we choose to turn on $\varphi_1$ and $\theta_1$ only, and turn off the rest of $\varphi_j$'s and $\theta_j$'s,
\begin{equation}
\varphi_1\,\ne\,0\,, \qquad \varphi_2\,=\,\varphi_3\,=\,\varphi_4\,=\,0\,.
\end{equation}
Then we only have to satisfy the $\theta_1$-equation in \eqref{thetas}: the scalar field, $\theta_1$, becomes a constant and the $\theta_1$-equation gives a condition,
\begin{equation} \label{firstcond}
q_1-q_2-q_3\,=\,0\,.
\end{equation}
Also from the equation, \eqref{precond}, we obtain another condition,
\begin{equation} \label{secondcond}
q_1-q_2+3q_3\,=\,\frac{2k}{g}\,.
\end{equation}
Then, from \eqref{firstcond} and \eqref{secondcond}, we obtain
\begin{equation} \label{qqq1}
q_1\,=\,\frac{1}{g}\left(-a+\frac{k}{4}\right)\,, \qquad q_2\,=\,\frac{1}{g}\left(-a-\frac{k}{4}\right)\,, \qquad q_3\,=\,\frac{k}{2g}\,,
\end{equation}
where $a$ is a constant. As we will see below, this case gives precisely the flow equations studied in \cite{Bobev:2014jva}. 

The flow equations reduce to
\begin{equation} 
\alpha'\,=\,\frac{g}{12}\frac{\partial{W}}{\partial\alpha}+X_{\alpha}\,,\qquad
\beta'\,=\,\frac{g}{4}\frac{\partial{W}}{\partial\beta}+X_{\beta}\,, \qquad 
\varphi_1'=\,\frac{g}{2}\frac{\partial{W}}{\partial\varphi_1}\,, \notag
\end{equation}
\begin{equation} 
U'\,=\,-\,\frac{g}{3}\,W\,+\,X_U\,, \qquad V'\,=\,-\,\frac{g}{3}\,W\,+\,X_V\,,
\end{equation}
and
\begin{align}
X_U\,=&\,-\,\frac{1}{3}\,e^{-2V}\,\left(q_1\,e^{2\alpha-2\beta}\,-\,q_2\,e^{2\alpha+2\beta}\,+\,q_3\,e^{-4\alpha}\right)\,, \notag \\
X_V\,=&\,+\,\frac{2}{3}\,e^{-2V}\,\left(q_1\,e^{2\alpha-2\beta}\,-\,q_2\,e^{2\alpha+2\beta}\,+\,q_3\,e^{-4\alpha}\right)\,, \notag \\
X_\alpha\,=&\,-\,\frac{1}{6}\,e^{-2V}\,\left(q_1\,e^{2\alpha-2\beta}\,-\,q_2\,e^{2\alpha+2\beta}\,-\,2\,q_3\,e^{-4\alpha}\right)\,, \notag \\
X_\beta\,=&\,+\,\frac{1}{2}\,e^{-2V}\,\left(q_1\,e^{2\alpha-2\beta}\,+\,q_2\,e^{2\alpha+2\beta}\right)\,,
\end{align}
where $q_1\,,\,q_2\,,\,q_3$ are given in \eqref{qqq1}, and the scalar fields, $\theta_j$, $j\,=\,1,\ldots, 4$, are constant. These flow equations indeed satisfy the equations of motion which are presented in appendix C.

We find $AdS_3\,\times\,\Sigma_k$ solutions of the flow equations. We set the scalar fields to be constant and
\begin{equation} \label{cb}
U(r)\,=\,U_0+\frac{r}{l}\,, \qquad V(r)\,=\,V_0\,,
\end{equation}
where $U_0$ and $V_0$ are constants. Then the flow equations are algebraic, and we obtain $AdS_3\,\times\,\Sigma_k$ solutions,
\begin{align}
e^{12\alpha}\,=\,\frac{4k^2}{k^2-16a^2}\,, \qquad e^{4\beta}\,&=\,\frac{-k-4a}{-k+4a}\,, \qquad e^{2\varphi_1}\,=\,\frac{-16a^2+5k^2-4k\sqrt{k^2-16a^2}}{16a^2+3k^2}\,, \notag \\
l^3\,=\,\frac{2}{k^2g^3}&(k^2-16a^2)\,, \qquad e^{6V}\,=\,\frac{1}{16g^6}\frac{(3k^2+16a^2)^3}{-k^3+16ka^2}\,,
\end{align}
where $|a|<1/4$. We see that this solution only makes sense when $k\,=\,-1$. After a proper normalization of the $r$-coordinate and taking $k\,=\,-1$, this is indeed the $AdS_3\,\times\,H_2$ solution found in (4.2) of \cite{Bobev:2014jva}, where the scalar field, $\varphi_1$, was denoted by $\chi$. Their flow equations look different from ours, but it is due to the different coordinates of the background metric from ours. In \cite{Bobev:2014jva}, the dual field theory of this solution was proposed to have 2d $\mathcal{N}\,=\,(0,2)$ supersymmetry, and central charge of the dual field theory was calculated and matched with the holographic calculation.

\subsubsection{Flow equations with two nontrivial scalar fields from hypermultiplets}

Now we consider turning on $\varphi_1$ and one of $\varphi_2$, $\varphi_3$, $\varphi_4$.

$\bullet$ First, we consider
\begin{equation}
\varphi_1\,\ne\,0\,, \qquad \varphi_2\,\ne\,0\,, \qquad \varphi_3\,=\,\varphi_4\,=\,0\,.
\end{equation}
To satisfy the equations, \eqref{thetas} and then \eqref{precond}, we should have
\begin{equation}
q_1-q_2-q_3\,=\,0\,, \qquad q_1+q_2+q_3\,=\,0\,, \qquad -q_2+q_3\,=\,\frac{k}{g}\,,
\end{equation}
and they give
\begin{equation} \label{qqq2}
q_1\,=\,0\,, \qquad q_2\,=\,-\frac{k}{2g}\,, \qquad q_3\,=\,\frac{k}{2g}\,.
\end{equation}
Hence, one of the gauge fields, $q_1$, is turned off, and there is no free parameter left like $a$ in the previous subsection.

The flow equations reduce to
\begin{equation} \label{nf1}
\alpha'\,=\,\frac{g}{12}\frac{\partial{W}}{\partial\alpha}+X_{\alpha}\,,\qquad
\beta'\,=\,\frac{g}{4}\frac{\partial{W}}{\partial\beta}+X_{\beta}\,, \qquad 
\varphi_1'=\,\frac{g}{2}\frac{\partial{W}}{\partial\varphi_1}\,, \qquad 
\varphi_2'=\,\frac{g}{2}\frac{\partial{W}}{\partial\varphi_2}\,, \notag
\end{equation}
\begin{equation}
U'\,=\,-\,\frac{g}{3}\,W\,+\,X_U\,, \qquad V'\,=\,-\,\frac{g}{3}\,W\,+\,X_V\,,
\end{equation}
and
\begin{align} \label{nf2}
X_U\,=&\,-\,\frac{1}{3}\,e^{-2V}\,\left(q_1\,e^{2\alpha-2\beta}\,-\,q_2\,e^{2\alpha+2\beta}\,+\,q_3\,e^{-4\alpha}\right)\,, \notag \\
X_V\,=&\,+\,\frac{2}{3}\,e^{-2V}\,\left(q_1\,e^{2\alpha-2\beta}\,-\,q_2\,e^{2\alpha+2\beta}\,+\,q_3\,e^{-4\alpha}\right)\,, \notag \\
X_\alpha\,=&\,-\,\frac{1}{6}\,e^{-2V}\,\left(q_1\,e^{2\alpha-2\beta}\,-\,q_2\,e^{2\alpha+2\beta}\,-\,2\,q_3\,e^{-4\alpha}\right)\,, \notag \\
X_\beta\,=&\,+\,\frac{1}{2}\,e^{-2V}\,\left(q_1\,e^{2\alpha-2\beta}\,+\,q_2\,e^{2\alpha+2\beta}\right)\,,
\end{align}
where $q_1\,,\,q_2\,,\,q_3$ are given in \eqref{qqq2}, and the scalar fields, $\theta_j$, $j\,=\,1,\ldots, 4$, are constant. These flow equations indeed satisfy the equations of motion. These flow equations are $new$ and our main result.

We find $AdS_3\,\times\,\Sigma_k$ solutions of the flow equations. We set the scalar fields to be constant and set as \eqref{cb} again. Then the flow equations are algebraic, and we obtain $AdS_3\,\times\,\Sigma_k$ solutions,
\begin{align} \label{ads32}
&e^{2\beta}\,=\,\frac{1}{2}\left(e^{6\alpha}+\sqrt{4+e^{12\alpha}}\right)\,, \qquad \varphi_1\,=\,0\,, \qquad \varphi_2\,=\,0\,, \notag \\ e^{2V_0}\,=&\,\frac{-ke^{-2\alpha}}{2g^2}\left(e^{6\alpha}+\sqrt{4+e^{12\alpha}}\right)\,, \qquad l\,=\,\frac{e^{2\alpha}}{g}\left(-e^{6\alpha}+\sqrt{4+e^{12\alpha}}\right)\,.
\end{align}
We see that this solution only makes sense when $k\,=\,-1$. This $AdS_3\,\times\,H_2$ solution is not new. This solution was first considered in section 3.2 of \cite{Maldacena:2000mw}, and recently discussed further in \cite{Benini:2013cda} and (4.5) of \cite{Bobev:2014jva}. The dual field theory of this solution was proposed to have 2d $\mathcal{N}\,=\,(2,2)$ supersymmetry, and central charge of the dual field theory was calculated and matched with the holographic calculation.

$\bullet$ Second, we consider
\begin{equation}
\varphi_1\,\ne\,0\,, \qquad \varphi_3\,\ne\,0\,, \qquad \varphi_2\,=\,\varphi_4\,=\,0\,.
\end{equation}
To satisfy the equations, \eqref{thetas} and then \eqref{precond}, we should have
\begin{equation}
q_1-q_2-q_3\,=\,0\,, \qquad -q_1-q_2+q_3\,=\,0\,, \qquad q_1+q_3\,=\,\frac{k}{g}\,,
\end{equation}
and they give
\begin{equation} \label{qqq3}
q_1\,=\,\frac{k}{2g}\,, \qquad q_2\,=\,0\,, \qquad q_3\,=\,\frac{k}{2g}\,.
\end{equation}
Hence, one of the gauge fields, $q_2$, is turned off, and there is no free parameter left.

The flow equations reduce to the identical form of \eqref{nf1} with \eqref{nf2}, but now $q_1$, $q_2$ and $q_3$ are given by \eqref{qqq3}. These flow equations also satisfy the equations of motion.

We can also set the scalar fields to be constant and the warp factors as in \eqref{cb}, and obtain $AdS_3\,\times\,\Sigma_k$ solutions,
\begin{align}
&e^{2\beta}\,=\,\frac{1}{2}\left(-e^{6\alpha}+\sqrt{4+e^{12\alpha}}\right)\,, \qquad \varphi_1\,=\,0\,, \qquad \varphi_2\,=\,0\,, \notag \\ e^{2V_0}\,=&\,\frac{-ke^{-2\alpha}}{2g^2}\left(e^{6\alpha}+\sqrt{4+e^{12\alpha}}\right)\,, \qquad l\,=\,\frac{e^{2\alpha}}{g}\left(-e^{6\alpha}+\sqrt{4+e^{12\alpha}}\right)\,.
\end{align}
This solution falls in the same class of the solution in \eqref{ads32}.

$\bullet$ Third, we consider
\begin{equation}
\varphi_1\,\ne\,0\,, \qquad \varphi_4\,\ne\,0\,, \qquad \varphi_2\,=\,\varphi_3\,=\,0\,.
\end{equation}
To satisfy the equations, \eqref{thetas} and then \eqref{precond}, we should have
\begin{equation}
q_1-q_2-q_3\,=\,0\,, \qquad q_1-q_2+q_3\,=\,0\,, \qquad q_3\,=\,\frac{k}{g}\,,
\end{equation}
but there is no $q_1$, $q_2$ and $q_3$ satisfying these conditions.

\subsubsection{Flow equations with three nontrivial scalar fields from hypermultiplets}

Lastly, we consider turning on $\varphi_1$ and two of $\varphi_2$, $\varphi_3$, $\varphi_4$. By observing the equations, \eqref{thetas} and \eqref{precond}, we find it not possible. 

We conclude that there could be {\it up to two} nontrivial scalar fields out of four, $\varphi_j$, $j\,=\,1,\ldots, 4$, with constant $\theta_j$, $j\,=\,1,\ldots, 4$, from hypermultiplets in our truncation.

\section{Conclusions}

In this paper, we obtained the magnetically-charged supersymmetric flow equations for scalar fields from two $\mathcal{N}$ = 2 vector multiplets and two $\mathcal{N}$ = 2 hypermultiplets by a consistent truncation of gauged $\mathcal{N}$ = 8 supergravity in five dimensions. We concluded that there could be only {\it up to two} nontrivial scalar fields out of eight from the two hypermultiplets in the flow equations. As we increase the number of nontrivial scalar fields from hypermultiplets, some $U(1)$ gauge fields had to be turned off. This should be related to the Higgs mechanism as scalar fields couple to gauge fields, gauge fields become massive.{\footnote{We would like to thank Krzysztof Pilch for comment on this.}} Along the way, we have rediscovered a number of known $AdS_3$ solutions. Our work could be understood as an extension of the flow equations presented in \cite{Bobev:2014jva} where only one scalar field from hypermultiplets were turned on.

Our flow equations provide a particular and concrete realization of the generic flow equations for general $\mathcal{N}$ = 2 matters in \cite{Klemm:2016kxw}. It would be interesting to further generalize our result to include more $\mathcal{N}$ = 2 matter multiplets by employing larger truncation of gauged $\mathcal{N}$ = 8 supergravity in five dimensions. Otherwise, it could be possible to rederive our results in the $\mathcal{N}$ = 2 supergravity formalism by employing and extending the scalar field parametrizations of \cite{Ceresole:2000jd, Ceresole:2001wi}.

\bigskip
\medskip
\leftline{\bf Acknowledgements}
We are very grateful to an anonymous referee who pointed out the contradiction in some of the supersymmetry equations. It resolved the inconsistency between the supersymmetry equations and the equations of motion in the first preprint of this paper. Conclusions were not changed. We would like to thank Krzysztof Pilch for helpful communications. We are grateful to Antoine Van Proeyen for explaining details of his work on $\mathcal{N}$ = 2 supergravity. This research was supported by the National Research Foundation of Korea under the grant NRF-2017R1D1A1B03034576.

\vspace{13cm}

\appendix
\section{Gauged $\mathcal{N}$ = 8 supergravity in five dimensions}
\renewcommand{\theequation}{A.\arabic{equation}}
\setcounter{equation}{0} 

In this appendix we review gauged $\mathcal{N}$ = 8 supergravity in five dimensions with emphasis on the structure of its scalar manifold, $E_{6(6)}/USp(8)$, by following \cite{Gunaydin:1985cu}. We will employ the conventions of \cite{Gunaydin:1985cu} throughout the paper.

The $SO(6)$ gauged $\mathcal{N}$ = 8 supergravity in five dimensions \cite{Pernici:1985ju, Gunaydin:1984qu, Gunaydin:1985cu} has local $USp(8)$ symmetry, but global $E_{6(6)}$ symmetry of the ungauged theory is broken. The field content consists of 1 graviton $e_{\mu}\,^{a}$, 8 gravitini $\psi_{\mu}\,^{a}$, 15 vector fields $A_{{\mu}IJ}$, 12 two-form tensor fields $B_{\mu\nu}\,^{I\alpha}$, 48 spinor fields $\chi^{abc}$, and 42 scalar fields $\phi^{abcd}$ where $a$, $b$, $\ldots$ are $USp(8)$ indices, $I$, $J$, $\ldots$ are $SL(6,\mathbb{R})$, and $\alpha$, $\beta$, $\ldots$ are $SL(2,\mathbb{R})$. Here $SL(6,\mathbb{R})$${\times}$$SL(2,\mathbb{R})$ is one of the maximal subgroups of $E_{6(6)}$.  

The infinitesimal $E_{6(6)}$ transformation in the $SL(6,\mathbb{R})$${\times}$$SL(2,\mathbb{R})$ basis, ($z_{IJ}$, $z^{I{\alpha}}$), in terms of $\Lambda^I\,_J$, $\Lambda^\alpha\,_\beta$, and $\Sigma_{IJK{\alpha}}$ was already given in \eqref{zls1} and \eqref{zls2}. Exponentiating the transformation in \eqref{zls1} and \eqref{zls2},
\begin{align} \label{zU1}
z'_{IJ}\,=\,&\,\,\frac{1}{2}\,U^{MN}\,_{IJ}\,z_{MN}\,+\,\sqrt{\frac{1}{2}}\,U_{P{\beta}IJ}\,z^{P{\beta}}\,, \\ \label{zU2}
z'^{K{\beta}}\,=\,&\,\,U_{P{\beta}}\,^{K{\alpha}}\,z^{P{\beta}}\,+\,\sqrt{\frac{1}{2}}\,U^{IJK{\alpha}}\,z_{IJ}\,,
\end{align}
we obtain the coset representatives in the $SL(6,\mathbb{R})$${\times}$$SL(2,\mathbb{R})$ basis, $U^{IJ}\,_{KL}$, $U^{IJK{\alpha}}$ and $U_{I{\alpha}}\,^{J{\beta}}$. We also have the coset representatives in the $USp(8)$ basis,
\begin{align} \label{UV1}
\mathcal{V}^{IJab}\,=\,&\,\,\frac{1}{8}\,\,\left[(\Gamma_{KL})^{ab}\,U^{IJ}\,_{KL}\,+\,2(\Gamma_{K{\beta}})^{ab}\,U^{IJK{\beta}}\right]\,, \\ \label{UV2}
\mathcal{V}_{I{\alpha}}\,^{ab}\,=\,&\,\,\frac{1}{4}\,\sqrt{\frac{1}{2}}\,\,\left[(\Gamma_{KL})^{ab}\,U_{I\alpha}\,^{KL}\,+\,2(\Gamma_{K{\beta}})^{ab}\,U_{I\alpha}\,^{K{\beta}}\right]\,.
\end{align}
The inverse coset representatives are
\begin{align}
\widetilde{\mathcal{V}}_{IJab}\,&=\,\frac{1}{8}\,[(\Gamma_{KL})_{ab}\,\widetilde{U}_{IJ}\,^{KL}\,+\,2\,(\Gamma_{K\alpha})_{ab}\,\widetilde{U}_{IJ}\,^{K\alpha}]\,, \\
\widetilde{\mathcal{V}}^{I\alpha}\,_{ab}\,&=\,\frac{1}{4}\,\sqrt{\frac{1}{2}}\,[(\Gamma_{KL})_{ab}\,\widetilde{U}^{I{\alpha}KL}\,+\,2\,(\Gamma_{K\beta})_{ab}\,\widetilde{U}_{I\alpha}\,^{K\beta}]\,.
\end{align}

Now we consider the action of the theory \cite{Gunaydin:1985cu}. The bosonic part of the Lagrangian is
\begin{equation} \label{N8lag}
e^{-1}\,\mathcal{L}\,=\,-\frac{1}{4}\,R\,+\,\mathcal{L}_{kin}\,+\,\mathcal{P}\,-\frac{1}{8}\,H_{{\mu}{\nu}{a}{b}}\,H^{{\mu}{\nu}{a}{b}}\,+\,\frac{1}{8\,g\,e}\,\epsilon^{\mu\nu\rho\sigma\tau}\,\epsilon_{\alpha\beta}\,B_{\mu\nu}\,^{I\alpha}\,D_{\rho}B_{\sigma\tau}\,^{I\beta}
\,+\,\mathcal{L}_{CS}\,,
\end{equation}
where the covariant derivative is defined by
\begin{equation}
D_{\mu}X_{aI}\,=\,\partial_{\mu}X_{aI}\,+\,Q_{{\mu}a}\,^{b}\,X_{bI}\,-\,g\,A_{{\mu}IJ}\,X_{aJ}\,,
\end{equation}
with the $USp(8)$ connection,
\begin{equation}
Q_{{\mu}a}\,^b\,=\,-\frac{1}{3}\,\left[\,\widetilde{\mathcal{V}}^{bcIJ}\,\partial_{\mu}\mathcal{V}_{IJac}\,+\,\widetilde{\mathcal{V}}^{bcI{\alpha}}\,{\partial}_{\mu}\mathcal{V}_{I{\alpha}ac}\,+\,g\,A_{{\mu}IL}\,\eta^{JL}\,(2\,V_{ae}\,^{IK}\,\widetilde{\mathcal{V}}^{be}\,_{JK}\,-\,\mathcal{V}_{J{\alpha}ae}\,\widetilde{\mathcal{V}}^{beI{\alpha}})\right]\,.
\end{equation}
The kinetic term for scalar fields is defined by
\begin{equation}
\mathcal{L}_{kin}\,=\,\,\frac{1}{24}\,P_{{\mu}abcd}\,P^{{\mu}abcd}\,,
\end{equation}
where
\begin{equation}
P_{\mu}\,^{abcd}\,=\,\widetilde{\mathcal{V}}^{ab}\,_{IJ}\,D_{\mu}\mathcal{V}^{IJcd}\,+\,\widetilde{\mathcal{V}}^{abI{\alpha}}\,D_{\mu}\mathcal{V}_{I{\alpha}}\,^{cd}\,.
\end{equation}
The scalar potential is defined by
\begin{equation}
\mathcal{P}\,=\,-\,\frac{1}{32}\,(2W_{ab}\,W^{ab}\,-\,W_{abcd}W^{abcd})\,,
\end{equation}
where
\begin{equation}
W_{abcd}\,=\,\epsilon^{\alpha\beta}\,\eta^{IJ}\,\mathcal{V}_{I{\alpha}ab}\,\mathcal{V}_{J{\beta}cd}\,,
\end{equation}
\begin{equation}
W_{ab}\,=\,W^c\,_{acb}\,.
\end{equation}
We also define
\begin{equation}
H_{\mu\nu}\,^{ab}\,=\,F_{\mu\nu}\,^{ab}+B_{\mu\nu}\,^{ab}\,, \end{equation}
where
\begin{align}
F_{\mu\nu}\,^{ab}\,&=\,F_{\mu\nu{IJ}}\,\mathcal{V}^{IJab}\,, \\
B_{\mu\nu}\,^{ab}\,&=\,B_{\mu\nu}\,^{I{\alpha}}\,\mathcal{V}_{I\alpha}\,^{ab}\,,
\end{align}
for the last three terms of Lagrangian.

We adopt the gamma matrix convention of \cite{Gunaydin:1985cu}, with
\begin{equation}
\{\gamma^{i},\,\gamma^{j}\}\,=\,2\,\eta^{ij}\,,
\end{equation}
where $\eta^{ij}\,=\,diag\,(+,\,-,\,-,\,-,\,-)$, and $\gamma^{0},\, \gamma^{1},\,\gamma^{2},\,\gamma^{3}$ are pure imaginary as in four-dimensions and $\gamma^{4}\,=\,i\,\gamma^{5}$ is pure real. The matrices $\gamma^{0}$ and $\gamma^{5}$ are antisymmetric and $\gamma^{1},\, \gamma^{2},\,\gamma^{3}$ are symmetric.

\section{The supersymmetry variations for spin-1/2 fields}
\renewcommand{\theequation}{B.\arabic{equation}}
\setcounter{equation}{0}

Some components of the spin-1/2 field variations in \eqref{1/2A} are
\begin{align}
&\delta\chi_{123}\,=\,\gamma^4\left[\frac{1}{4}(3\alpha'-\beta')\hat{\epsilon}_1+i\frac{1}{4}(\sin\theta_2\varphi_2'+\sin\theta_4\varphi_4')\hat{\epsilon}_1-i\frac{1}{4}(\cos\theta_2\varphi_2'+\cos\theta_4\varphi_4')\hat{\epsilon}_2\right] \notag \\
+\gamma^2&\left[i\frac{g}{8}\left(\cos\theta_2\sinh(2\varphi_2)(\dot{\theta}_2-(q_1+q_2+q_3)\int{f}_kdx)+\cos\theta_4\sinh(2\varphi_4)(\dot{\theta}_4-(q_1-q_2+q_3)\int{f}_kdx)\right)\hat{\epsilon}_1\right. \notag \\
+&\left.i\frac{g}{8}\left(\sin\theta_2\sinh(2\varphi_2)(\dot{\theta}_2-(q_1+q_2+q_3)\int{f}_kdx)+\sin\theta_4\sinh(2\varphi_4)(\dot{\theta}_4-(q_1-q_2+q_3)\int{f}_kdx)\right)\hat{\epsilon}_2\right] \notag \\
-&\frac{g}{2}\left[\frac{1}{8}\left(e^{-2\alpha+2\beta}(\cosh(2\varphi_1)+\cosh(2\varphi_2)-\cosh(2\varphi_3)+\cosh(2\varphi_4))\right.\right. \notag \\
&\left.+e^{4\alpha}\left(\cosh(2\varphi_1)-\cosh(2\varphi_2)-\cosh(2\varphi_3)-\cosh(2\varphi_4)\right)\right)\hat{\epsilon}_2 \notag \\
-i&\frac{1}{8}\left((e^{-2\alpha+2\beta}-e^{-2\alpha-2\beta}+e^{4\alpha})\cos\theta_2\sinh(2\varphi_2)+(e^{-2\alpha+2\beta}+e^{-2\alpha-2\beta}+e^{4\alpha})\cos\theta_4\sinh(2\varphi_4)\right)\hat{\epsilon}_1 \notag \\
-i&\left.\frac{1}{8}\left((e^{-2\alpha+2\beta}-e^{-2\alpha-2\beta}+e^{4\alpha})\sin\theta_2\sinh(2\varphi_2)+(e^{-2\alpha+2\beta}+e^{-2\alpha-2\beta}+e^{4\alpha})\sin\theta_4\sinh(2\varphi_4)\right)\hat{\epsilon}_2\right] \notag \\
+&\frac{1}{4}e^{-2V}\gamma^1\gamma^2(-e^{2\alpha-2\beta}q_1+e^{-4\alpha}q_3)\hat{\epsilon}_1\,=\,0\,,
\end{align}
\begin{align}
&\delta\chi_{456}\,=\,\gamma^4\left[-\frac{1}{4}(3\alpha'+\beta')\hat{\epsilon}_2+i\frac{1}{4}(\sin\theta_3\varphi_3'-\sin\theta_4\varphi_4')\hat{\epsilon}_2-i\frac{1}{4}(\cos\theta_3\varphi_3'+\cos\theta_4\varphi_4')\hat{\epsilon}_1\right] \notag \\
+\gamma^2&\left[i\frac{g}{8}\left(\cos\theta_3\sinh(2\varphi_3)(\dot{\theta}_3+(-q_1-q_2+q_3)\int{f}_kdx)-\cos\theta_4\sinh(2\varphi_4)(\dot{\theta}_4-(q_1-q_2+q_3)\int{f}_kdx)\right)\hat{\epsilon}_2\right. \notag \\
+&\left.i\frac{g}{8}\left(\sin\theta_3\sinh(2\varphi_3)(\dot{\theta}_3+(-q_1-q_2+q_3)\int{f}_kdx)+\sin\theta_4\sinh(2\varphi_4)(\dot{\theta}_4-(q_1-q_2+q_3)\int{f}_kdx)\right)\hat{\epsilon}_1\right] \notag \\
-&\frac{g}{2}\left[\frac{1}{8}\left(e^{-2\alpha-2\beta}(\cosh(2\varphi_1)-\cosh(2\varphi_2)+\cosh(2\varphi_3)+\cosh(2\varphi_4))\right.\right. \notag \\
&\left.+e^{4\alpha}\left(\cosh(2\varphi_1)-\cosh(2\varphi_2)-\cosh(2\varphi_3)-\cosh(2\varphi_4)\right)\right)\hat{\epsilon}_1 \notag \\
-i&\frac{1}{8}\left((e^{-2\alpha+2\beta}-e^{-2\alpha-2\beta}-e^{4\alpha})\cos\theta_3\sinh(2\varphi_3)-(e^{-2\alpha+2\beta}+e^{-2\alpha-2\beta}+e^{4\alpha})\cos\theta_4\sinh(2\varphi_4)\right)\hat{\epsilon}_2 \notag \\
-i&\left.\frac{1}{8}\left((e^{-2\alpha+2\beta}-e^{-2\alpha-2\beta}-e^{4\alpha})\sin\theta_3\sinh(2\varphi_3)+(e^{-2\alpha+2\beta}+e^{-2\alpha-2\beta}+e^{4\alpha})\sin\theta_4\sinh(2\varphi_4)\right)\hat{\epsilon}_1\right] \notag \\
-&\frac{1}{4}e^{-2V}\gamma^1\gamma^2(e^{2\alpha+2\beta}q_2+e^{-4\alpha}q_3)\hat{\epsilon}_2\,=\,0\,.
\end{align}

\section{The equations of motion}
\renewcommand{\theequation}{C.\arabic{equation}}
\setcounter{equation}{0}

In this appendix, we give the equations of motion of the consistent truncation in \eqref{lag}. The Einstein equations are
\begin{align}
R_{\mu\nu}&-\frac{1}{2}Rg_{\mu\nu}-2\mathcal{P}g_{\mu\nu}+2\left(e^{4\alpha-4\beta}T^{A_1}_{\mu\nu}+e^{4\alpha+4\beta}T^{A_2}_{\mu\nu}+e^{-8\alpha}T^{A_3}_{\mu\nu}\right) \notag \\
&-2\left(6T^\alpha_{\mu\nu}+2T^\beta_{\mu\nu}\right)-2\left(T^{\varphi_1}_{\mu\nu}+T^{\varphi_2}_{\mu\nu}+T^{\varphi_3}_{\mu\nu}+T^{\varphi_4}_{\mu\nu}\right) \notag \\
&-2\left(\frac{1}{4}\sinh^2(2\varphi_1)T^{\theta_1}_{\mu\nu}+\frac{1}{4}\sinh^2(2\varphi_2)T^{\theta_2}_{\mu\nu}+\frac{1}{4}\sinh^2(2\varphi_3)T^{\theta_3}_{\mu\nu}+\frac{1}{4}\sinh^2(2\varphi_4)T^{\theta_4}_{\mu\nu}\right)\,=\,0,
\end{align}
and
\begin{align}
T^{A_i}_{\mu\nu}\,=\,g^{\rho\sigma}F^{(i)}_{\mu\rho}F^{(i)}_{\nu\sigma}-\frac{1}{4}g_{\mu\nu}F^{(i)}_{\rho\sigma}F^{(i)\rho\sigma} \notag \\
T^{X}_{\mu\nu}\,=\,\partial_\mu{X}\partial_\nu{X}-\frac{1}{2}g_{\mu\nu}\partial_\rho{X}\partial^\rho{X}\,,
\end{align}
where $X$ denotes a scalar field. The Maxwell equations are
\begin{align} \label{maxwell}
&\partial_\nu\left(\sqrt{g}e^{4\alpha-4\beta}F^{(1)}\,_\mu\,^\nu\right) \notag \\
&-\frac{1}{4}\sinh^2(2\varphi_1)\left(\partial_\mu\theta_1+(A^{(1)}_\mu-A^{(2)}_\mu-A^{(3)}_\mu)\right)-\frac{1}{4}\sinh^2(2\varphi_2)\left(\partial_\mu\theta_2-(A^{(1)}_\mu+A^{(2)}_\mu+A^{(3)}_\mu)\right) \notag \\ 
&-\frac{1}{4}\sinh^2(2\varphi_3)\left(\partial_\mu\theta_3+(-A^{(1)}_\mu-A^{(2)}_\mu+A^{(3)}_\mu)\right)-\frac{1}{4}\sinh^2(2\varphi_4)\left(\partial_\mu\theta_4-(A^{(1)}_\mu-A^{(2)}_\mu+A^{(3)}_\mu)\right)\,=\,0\,, \notag
\end{align}
\begin{align}
&\partial_\nu\left(\sqrt{g}e^{4\alpha+4\beta}F^{(2)}\,_\mu\,^\nu\right) \notag \\
&-\frac{1}{4}\sinh^2(2\varphi_1)\left(\partial_\mu\theta_1+(A^{(1)}_\mu-A^{(2)}_\mu-A^{(3)}_\mu)\right)-\frac{1}{4}\sinh^2(2\varphi_2)\left(\partial_\mu\theta_2-(A^{(1)}_\mu+A^{(2)}_\mu+A^{(3)}_\mu)\right) \notag \\ 
&-\frac{1}{4}\sinh^2(2\varphi_3)\left(\partial_\mu\theta_3+(-A^{(1)}_\mu-A^{(2)}_\mu+A^{(3)}_\mu)\right)-\frac{1}{4}\sinh^2(2\varphi_4)\left(\partial_\mu\theta_4-(A^{(1)}_\mu-A^{(2)}_\mu+A^{(3)}_\mu)\right)\,=\,0\,, \notag 
\end{align}
\begin{align}
&\partial_\nu\left(\sqrt{g}e^{-8\alpha}F^{(3)}\,_\mu\,^\nu\right) \notag \\
&-\frac{1}{4}\sinh^2(2\varphi_1)\left(\partial_\mu\theta_1+(A^{(1)}_\mu-A^{(2)}_\mu-A^{(3)}_\mu)\right)-\frac{1}{4}\sinh^2(2\varphi_2)\left(\partial_\mu\theta_2-(A^{(1)}_\mu+A^{(2)}_\mu+A^{(3)}_\mu)\right) \notag \\ 
&-\frac{1}{4}\sinh^2(2\varphi_3)\left(\partial_\mu\theta_3+(-A^{(1)}_\mu-A^{(2)}_\mu+A^{(3)}_\mu)\right)-\frac{1}{4}\sinh^2(2\varphi_4)\left(\partial_\mu\theta_4-(A^{(1)}_\mu-A^{(2)}_\mu+A^{(3)}_\mu)\right)\,=\,0\,.
\end{align}
The scalar field equations are
\begin{align} \label{eomalbe}
&\frac{1}{\sqrt{g}}\partial_\mu\left(\sqrt{g}g^{\mu\nu}\partial_\nu\alpha\right)+\frac{1}{6}\left(e^{4\alpha-4\beta}F^{(1)}_{\mu\nu}F^{(1)\mu\nu}+e^{4\alpha+4\beta}F^{(2)}_{\mu\nu}F^{(2)\mu\nu}-2e^{-8\alpha}F^{(3)}_{\mu\nu}F^{(3)\mu\nu}\right)+\frac{1}{6}\frac{\partial\mathcal{P}}{\partial\alpha}\,=\,0\,, \notag \\
&\frac{1}{\sqrt{g}}\partial_\mu\left(\sqrt{g}g^{\mu\nu}\partial_\nu\beta\right)+\frac{1}{2}\left(-e^{4\alpha-4\beta}F^{(1)}_{\mu\nu}F^{(1)\mu\nu}+e^{4\alpha+4\beta}F^{(2)}_{\mu\nu}F^{(2)\mu\nu}\right)+\frac{1}{2}\frac{\partial\mathcal{P}}{\partial\beta}\,=\,0\,,
\end{align}
\begin{align} \label{eomvarphi}
\frac{1}{\sqrt{g}}\partial_\mu\left(\sqrt{g}g^{\mu\nu}\partial_\nu\varphi_1\right)+\frac{1}{4}\sinh(4\varphi_1)\left(\partial_\mu\theta_1+(A^{(1)}_\mu-A^{(2)}_\mu-A^{(3)}_\mu)\right)^2+\frac{\partial\mathcal{P}}{\partial\varphi_1}\,=\,0\,, \notag \\
\frac{1}{\sqrt{g}}\partial_\mu\left(\sqrt{g}g^{\mu\nu}\partial_\nu\varphi_2\right)+\frac{1}{4}\sinh(4\varphi_2)\left(\partial_\mu\theta_2-(A^{(1)}_\mu+A^{(2)}_\mu+A^{(3)}_\mu)\right)^2+\frac{\partial\mathcal{P}}{\partial\varphi_2}\,=\,0\,, \notag \\
\frac{1}{\sqrt{g}}\partial_\mu\left(\sqrt{g}g^{\mu\nu}\partial_\nu\varphi_3\right)+\frac{1}{4}\sinh(4\varphi_3)\left(\partial_\mu\theta_3+(-A^{(1)}_\mu-A^{(2)}_\mu+A^{(3)}_\mu)\right)^2+\frac{\partial\mathcal{P}}{\partial\varphi_3}\,=\,0\,, \notag \\
\frac{1}{\sqrt{g}}\partial_\mu\left(\sqrt{g}g^{\mu\nu}\partial_\nu\varphi_4\right)+\frac{1}{4}\sinh(4\varphi_4)\left(\partial_\mu\theta_4-(A^{(1)}_\mu-A^{(2)}_\mu+A^{(3)}_\mu)\right)^2+\frac{\partial\mathcal{P}}{\partial\varphi_4}\,=\,0\,.
\end{align}
and
\begin{align} \label{eomtheta}
\partial_\nu\left(\sqrt{g}g^{\mu\nu}\sinh^2(2\varphi_1)(\partial_\mu\theta_1+(A^{(1)}_\mu-A^{(2)}_\mu-A^{(3)}_\mu))\right)\,=\,0\,, \notag \\
\partial_\nu\left(\sqrt{g}g^{\mu\nu}\sinh^2(2\varphi_2)(\partial_\mu\theta_2-(A^{(1)}_\mu+A^{(2)}_\mu+A^{(3)}_\mu))\right)\,=\,0\,, \notag \\
\partial_\nu\left(\sqrt{g}g^{\mu\nu}\sinh^2(2\varphi_3)(\partial_\mu\theta_3+(-A^{(1)}_\mu-A^{(2)}_\mu+A^{(3)}_\mu))\right)\,=\,0\,, \notag \\
\partial_\nu\left(\sqrt{g}g^{\mu\nu}\sinh^2(2\varphi_4)(\partial_\mu\theta_4-(A^{(1)}_\mu-A^{(2)}_\mu+A^{(3)}_\mu))\right)\,=\,0\,.
\end{align}

\vspace{12cm}




\begin{thebibliography}{99}

\bibitem{Maldacena:1997re}
  J.~M.~Maldacena,
  {\it The large N limit of superconformal field theories and supergravity,}
  Adv.\ Theor.\ Math.\ Phys.\  {\bf 2}, 231 (1998)
  [Int.\ J.\ Theor.\ Phys.\  {\bf 38}, 1113 (1999)]
  [arXiv:hep-th/9711200].

\bibitem{Chamseddine:1999xk} 
  A.~H.~Chamseddine and W.~A.~Sabra,
  {\it Magnetic strings in five-dimensional gauged supergravity theories,}  Phys.\ Lett.\ B {\bf 477}, 329 (2000)  [hep-th/9911195].

\bibitem{Klemm:2000nj} 
  D.~Klemm and W.~A.~Sabra,
  {\it Supersymmetry of black strings in D = 5 gauged supergravities,}  Phys.\ Rev.\ D {\bf 62}, 024003 (2000)  [hep-th/0001131].

\bibitem{Cacciatori:2003kv} 
  S.~L.~Cacciatori, D.~Klemm and W.~A.~Sabra,
  {\it Supersymmetric domain walls and strings in D = 5 gauged supergravity coupled to vector multiplets,}  JHEP {\bf 0303}, 023 (2003)  [hep-th/0302218].

\bibitem{D'Hoker:2009mm} 
  E.~D'Hoker and P.~Kraus,
  {\it Magnetic Brane Solutions in AdS,}  JHEP {\bf 0910}, 088 (2009)  [arXiv:0908.3875 [hep-th]].

\bibitem{Almuhairi:2010rb} 
  A.~Almuhairi,
  {\it AdS$_3$ and AdS$_2$ Magnetic Brane Solutions,}  arXiv:1011.1266 [hep-th].

\bibitem{Almuhairi:2011ws} 
  A.~Almuhairi and J.~Polchinski,
  {\it Magnetic AdS x R$^2$: Supersymmetry and stability,}  arXiv:1108.1213 [hep-th].

\bibitem{Donos:2011pn} 
  A.~Donos, J.~P.~Gauntlett and C.~Pantelidou,
  {\it Magnetic and Electric AdS Solutions in String- and M-Theory,}  Class.\ Quant.\ Grav.\  {\bf 29}, 194006 (2012)  [arXiv:1112.4195 [hep-th]].

\bibitem{Maldacena:2000mw} 
  J.~M.~Maldacena and C.~Nunez,
  {\it Supergravity description of field theories on curved manifolds and a no go theorem,}
  Int.\ J.\ Mod.\ Phys.\ A {\bf 16}, 822 (2001) [hep-th/0007018].

\bibitem{Naka:2002jz} 
  M.~Naka,
  {\it Various wrapped branes from gauged supergravities,}
  hep-th/0206141.

\bibitem{Cucu:2003bm} 
  S.~Cucu, H.~Lu and J.~F.~Vazquez-Poritz,
  {\it A Supersymmetric and smooth compactification of M theory to AdS(5),}
  Phys.\ Lett.\ B {\bf 568}, 261 (2003) [hep-th/0303211].

\bibitem{Cucu:2003yk} 
  S.~Cucu, H.~Lu and J.~F.~Vazquez-Poritz,
  {\it Interpolating from AdS(D-2) x S**2 to AdS(D),}
  Nucl.\ Phys.\ B {\bf 677}, 181 (2004) [hep-th/0304022].

\bibitem{Benini:2012cz} 
  F.~Benini and N.~Bobev,
  {\it Exact two-dimensional superconformal R-symmetry and c-extremization,}
  Phys.\ Rev.\ Lett.\  {\bf 110}, no. 6, 061601 (2013) [arXiv:1211.4030 [hep-th]].

\bibitem{Benini:2013cda} 
  F.~Benini and N.~Bobev,
  {\it Two-dimensional SCFTs from wrapped branes and c-extremization,}
  JHEP {\bf 1306}, 005 (2013) [arXiv:1302.4451 [hep-th]].

\bibitem{Karndumri:2013iqa} 
  P.~Karndumri and E.~O Colgain,
  {\it Supergravity dual of $c$-extremization,}
  Phys.\ Rev.\ D {\bf 87}, no. 10, 101902 (2013) [arXiv:1302.6532 [hep-th]].

\bibitem{Amariti:2016mnz} 
  A.~Amariti and C.~Toldo,
  {\it Betti multiplets, flows across dimensions and c-extremization,}
  JHEP {\bf 1707}, 040 (2017) [arXiv:1610.08858 [hep-th]].

\bibitem{Benini:2015bwz} 
  F.~Benini, N.~Bobev and P.~M.~Crichigno,
  {\it Two-dimensional SCFTs from D3-branes,}
  JHEP {\bf 1607}, 020 (2016) [arXiv:1511.09462 [hep-th]].

\bibitem{Bobev:2014jva} 
  N.~Bobev, K.~Pilch and O.~Vasilakis,
  {\it (0, 2) SCFTs from the Leigh-Strassler fixed point,}
  JHEP {\bf 1406}, 094 (2014) [arXiv:1403.7131 [hep-th]].

\bibitem{Klemm:2016kxw} 
  D.~Klemm, N.~Petri and M.~Rabbiosi,
  {\it Black string first order flow in $N = 2, d = 5$ abelian gauged supergravity,}
  JHEP {\bf 1701}, 106 (2017) [arXiv:1610.07367 [hep-th]].

\bibitem{Khavaev:2000gb} 
  A.~Khavaev and N.~P.~Warner,
  {\it A Class of N=1 supersymmetric RG flows from five-dimensional N=8 supergravity,}  Phys.\ Lett.\ B {\bf 495}, 215 (2000)  [hep-th/0009159].

\bibitem{Bobev:2010de}
  N.~Bobev, A.~Kundu, K.~Pilch and N.~P.~Warner,
  {\it Supersymmetric Charged Clouds in $AdS_5$,}
  JHEP {\bf 1103}, 070 (2011)
  [arXiv:1005.3552 [hep-th]].

\bibitem{Pernici:1985ju}
  M.~Pernici, K.~Pilch and P.~van Nieuwenhuizen,
  {\it Gauged N=8 D=5 Supergravity,}
  Nucl.\ Phys.\  B {\bf 259}, 460 (1985).

\bibitem{Gunaydin:1984qu}
  M.~Gunaydin, L.~J.~Romans and N.~P.~Warner,
  {\it Gauged N=8 Supergravity In Five-Dimensions,}
  Phys.\ Lett.\  B {\bf 154}, 268 (1985).

\bibitem{Gunaydin:1985cu}
  M.~Gunaydin, L.~J.~Romans and N.~P.~Warner,
  {\it Compact And Noncompact Gauged Supergravity Theories In Five-Dimensions,}
  Nucl.\ Phys.\  B {\bf 272}, 598 (1986).

\bibitem{Freedman:1999gp} 
  D.~Z.~Freedman, S.~S.~Gubser, K.~Pilch and N.~P.~Warner,
  {\it Renormalization group flows from holography supersymmetry and a c theorem,}
  Adv.\ Theor.\ Math.\ Phys.\  {\bf 3}, 363 (1999) [hep-th/9904017].

\bibitem{Ceresole:2000jd} 
  A.~Ceresole and G.~Dall'Agata,
  {\it General matter coupled N=2, D = 5 gauged supergravity,}
  Nucl.\ Phys.\ B {\bf 585}, 143 (2000) [hep-th/0004111].

\bibitem{Ceresole:2001wi} 
  A.~Ceresole, G.~Dall'Agata, R.~Kallosh and A.~Van Proeyen,
  {\it Hypermultiplets, domain walls and supersymmetric attractors,}
  Phys.\ Rev.\ D {\bf 64}, 104006 (2001) [hep-th/0104056].

\end{thebibliography}
\end{document}